# MgB$_2$ thick film with $T_C$ = 40.2 K deposited on sapphire substrate

**Kaicheng Zhang, Li-li Ding, Cheng-gang Zhuang, Li-ping Chen, Chinping. Chen*, and Qing-rong Feng**

Department of physics, Peking university, Beijing 100871, China

We have successfully deposited thick MgB$_2$ film on the (0001) crystalline surface of sapphire by the method of hybrid physical-chemical vapor deposition (HPCVD). The film thickness is about 1.3 μm. It has a dense and interlaced structure. The film surface, shown by SEM, is stacked with MgB$_2$ microcrystals. Transport measurements by the 4-probe technique have demonstrated that its critical temperature is about 40.2 K, with a sharp transition width of 0.15 K. The residual resistivity ratio (RRR) is about 11. By extrapolation, $H_{C2}(0)$ is determined as 13.7 T from the magneto-transport measurement. Also by hysteresis measurement and applying the Bean model, the critical current density is estimated as $5 \times 10^{10}$ A/m$^2$ in zero magnetic field. The present work has demonstrated that HPCVD is an effective technique to fabricate the MgB$_2$ thick film with decent superconducting properties. Hence, it is important for the future superconducting application, in particular, as a crucial preliminary stage to fabricate superconducting tape.

## 1    Introduction

Since the superconductivity of MgB$_2$ was discovered [1], researchers have put great efforts in studying the properties of this material, from theoretic to experimental aspects. Theoretically, it is clear that MgB$_2$ is a BCS superconductor with the electron-phonon coupling mechanism, different from that of high-$T_C$ copper oxides. Experimentally, MgB$_2$ is attractive due to its double energy gaps. One is due to the hole-pairing in σ bond, the other, π bond. For the superconduction, σ band is dominant and the inharmonic effect of electron-phonon coupling have much contribution to it. Another interesting point with MgB$_2$ is that the critical temperature near 39 K is well beyond the prediction of BCS theory. This makes MgB$_2$ a special superconducting material for the study on its physical properties.

In the aspect of applied technology, MgB$_2$ has a great potential to supplant the widely used NbTi and Nb$_3$Sn superconducting wire in the superconducting magnet. It has the advantages of high transition temperature, in comparison with the two above-mentioned materials, however, without the disadvantages of the high-$T_C$ copper oxides. The advantages are as follow. Firstly, the coherence length

---

* Corresponding author: e-mail: cpchen@pku.edu.cn, Phone: +86   10   6275 1751, Fax: +86 10 6275 1615

of $MgB_2$ is long, ~ 5 nm, and the crystalline boundary has little effect on superconducting current. Also due to its good microstructure, $MgB_2$ is easy to achieve the higher critical superconducting current density as well as the stronger pinning state. Secondly, $MgB_2$ has the surface and crystalline boundary properties appropriate to form a superconducting junction. Therefore, it has a promising potential to achieve a good superconducting connection between two $MgB_2$ superconductors. This is very important for the superconducting magnet to operate in the persistent mode. Thirdly, $MgB_2$ can have much greater upper critical field and lower anisotropy with the doping. A little amount of carbon impurity can greatly improve the upper critical field of $MgB_2$ without suppressing its $T_C$ too much. It has been demonstrated by A.V. Pogrebnyakov et al. with the carbon-doped, epitaxial thin film that the upper critical field reaches as high as 70 T [2]. This greatly exceeds the value of current Nb-based superconducting material. Also the highest superconducting current density, $J_C$, of $MgB_2$ thin film has been reported to reach $4.0 \times 10^{11}$ $A/m^2$ [3], exceeding those of NbTi and $Nb_3Sn$ in the same condition. With the wire made by the method of powder-in-tube aiming for practical application, $J_C$, also achieves the value above $10^9$ $A/m^2$ [4].

It is well known that films deposited on different substrates have different critical temperatures. The highest one was recorded on the SiC substrate by the fabrication technique of HPCVD. It shows $T_C$ ~ 42 K. This exceeds the value of 39 K in most of the samples reported. Films deposited on sapphire also demonstrate a shift in $T_C$ to 40 K [5]. According to A.V. Pogrebnyakov et al. [5], the epitaxial films deposited on sapphire and SiC substrate with $T_C$ higher than 39 K are attributed to the strain in the film. The strain changes the electron-phonon coupling and results in the up shift of $T_C$. The film in the above investigation is epitaxial and is thin with the thickness ranging from hundreds to thousands of angstroms. Although HPCVD has been demonstrated as an effective method to fabricate high quality $MgB_2$ thin film, thick polycrystalline film deposited on sapphire with comparatively good superconducting properties has never been fabricated by the same method. We have successfully deposited thick $MgB_2$ film, about 1.3 μm, on the (0001) crystalline surface of $Al_2O_3$. The film exhibits $T_C$ of 40.2 K with a very narrow transition width, $\Delta T$ ~ 0.15 K. This is interesting with the up shift of $T_C$ from the usually reported 39 K for such material. In addition, a high quality thick film is important as a pre-stage for the purpose of superconducting application, in particular, in making a practical superconducting tape. In this article, we report the fabrication of $MgB_2$ thick film along with the characterizations and measurements on the film. Although, similar thick $MgB_2$ film on the sapphire substrate synthesized by the same technique has been reported previously by our group [6], the film property in this report has shown a significant improvement from the previous one. In particular, the interesting up shift of $T_C$ exceeding 40 K has been observed with the present thick film. An important underlying mechanism may have existed other than the internal strain suggested by A.V. Pogrebnyakov et al. for this property[5], since the present film is formed of microcrystallites without the possibility to generate the internal strain.

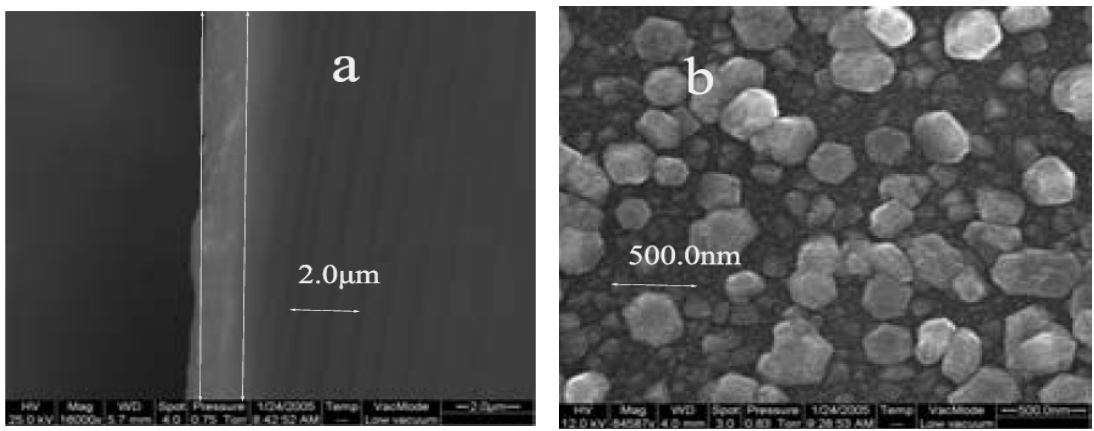

**Fig. 1** SEM images of MgB$_2$ film deposited on sapphire. (a) A view on the cross section of the film along with the substrate. The film thickness, as shown with the lighter shade in the middle between the two vertical lines, is about 1.3 μm. (b) A view on the film surface. It is covered with hexangular crystallites with diameter of 200 nm.

## 2    Sample preparation and characterization

The technique to fabricate the MgB$_2$ thick film in the present work is by the HPCVD, the same as that in our previous report [6], however, with different conditions. Bulk Mg (99.99%) was put on the platform inside the quartz pipe along with the substrate of Al$_2$O$_3$, which was about 10 mm×3 mm×1 mm with a surface of (0001) plane. Then, the quartz pipe was sealed and pumped down to the level of 20 ~ 30 Pa. Afterwards, the hydrogen gas (99.99%) was admitted into the pipe and regulated at the pressure of 20 kPa. The heater was switched on subsequently to heat the platform to 745 $^0$C. The substrate and the sample holder was sitting on the platform. The temperature and the pressure were kept constant with the admission of B$_2$H$_6$. The flowing rate of B$_2$H$_6$ and H$_2$ was maintained at $10^{-5}$ m$^3$/s and 2.2×$10^{-4}$ m$^3$/s, respectively, at 745 $^0$C for ten minutes. Then, the B$_2$H$_6$ was valved off and the heater, swithed off. The whole furnace was then staying in the atmosphere of H$_2$ until the furnace cooled down to room temperature. The film thus-obtained was shining with a purple color.

The surface morphology and the film thickness were examined by SEM, shown in Fig. 1. Fig 1(a) shows the cross sectional image of the film with the substrate. The film thickness shown between the two vertical lines in the middle is estimated about 1.3 μm. With the image of the film surface, shown in Fig. 1(b), it is apparent that the film is densely stacked with hexangular crystallites of MgB$_2$ approximately 200 nm in size.

## 3    Measurements and discussion

The electrical transport measurement was carried out on Quantum Design PPMS-9 by the 4-probe technique, using gold wire as the electrical leads and silver paste as the adhesive. The contact resistance is less than 1Ω. For the transport measurement, the current is selected as 200 μA in dc mode and the cooling rate is 2 K/min between 300 K and 50 K, 0.2 K/min between 50 K and 6 K. Also, for the magneto-transport investigation, the applied magnetic field is perpendicular to the film with the field strength of 0.5 T, 1 T, 3 T, 5 T, 7 T, and 8.5 T. The temperature-dependent resistivity is shown in Fig. 2(a), while the magneto-resistivity, in Fig. 2(b).

The $T_C$(onset), determined from Fig. 2(a), is 40.2 K with the transition width $\Delta T_C$ roughly equal to 0.15 K, shown in the inset. In the normal state, the resistivity is 0.267 μΩ·m at room temperature and 0.024 μΩ·m, slightly above the transition temperature of 40.2 K. This value falls within the resistivity range for a clean MgB$_2$ film at 40 K, from $10^{-3}$ to $10^{-1}$ μΩ·m [7]. The residual resistivity ratio, RRR = $\rho(300K)/\rho(40.2K)$, is 11, which is very high comparing with the samples reported in many previous experiments. It implies that the film is in the clean limit. To estimate the electron mean free path at 40 K, we use, $\ell = mv_F / ne^2 \rho_{40K}$, where $m$ is the mass of electron. The film resistivity at $T$ = 40.2 K is $\rho_{40K}$ = 0.024 μΩ·m determined from the measurement. For the estimation, we use the Fermi velocity [8], $v_F$ = 4.8×$10^5$ m/s, and the carrier density, n = 6.7×$10^{28}$ e/m$^3$ (two free electron per unit cell). The mean-free path of electron at 40.2 K is then estimated as 10 nm, which is larger than the coherence length of 5 nm [9]. This also satisfies the clean limit of superconductor. It is noteworthy that the mean free path, ~ 10 nm, is much smaller than the size of MgB$_2$ crystallites, ~ 200 nm. This implies that the estimation based on the Drude model is a reasonable one. Also, the value of resistivity calculated from the present sample is close to the value of single crystal [10]. This indicates that the effect of boundary scattering is small. .

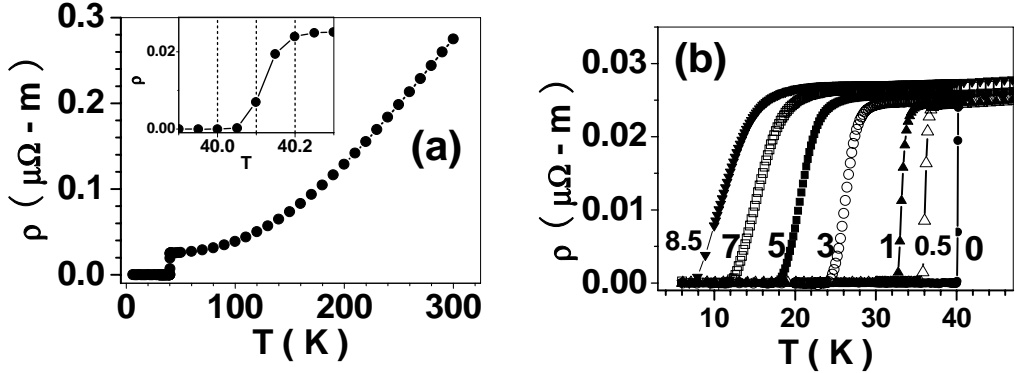

**Fig. 2** Temperature-dependent resistivity measurements. (a) ρ–T curve of the film under zero applied field. The inset shows the behavior near $T_C$. The $T_C$ onset is at 40.2 K and the transition width $\Delta T_C$ is roughly equal to 0.15 K. (b) R-T curves obtained under different applied fields, including 0, 0.5, 1, 3, 5, 7, and 8.5 T.

In comparison with the resistivity of $Nb_3Sn$ in the normal state at 20 K, ~ 0.11 μΩ·m, and at 300 K, ~ 0.80 μΩ·m [11], the values of the present sample at 40 K and 300 K are smaller by a factor of 3 to 4. It suggests a highly competitive application potential with $MgB_2$. We fit the curve of the temperature-dependent resistivity from 40 K to 200 K by the power law, $\rho = \rho_0 + \rho_1 T^\alpha$. The exponent, α, is obtained as 2.05. This value is lower than 2.6 with the $MgB_2$ wire [12] and 2.8 with the sintered $MgB_2$ sample [9]. The resistivity arising from the phonon scattering is described by the following function [13], $\rho(T) = \rho_0 [J_5(\Theta_D/T)]/(\Theta_D/T)^5$, with $J_5(Y) = \int_0^Y x^5/(e^x-1)(1-e^{-x})dx$. In the above expression, $\Theta_D$ is the Debye temperature, close to 800 K for $MgB_2$ [14], $\rho_0$ is a constant depending on the material. The data generated numerically using the above function in the temperature range from 40 K to 200 K can be well fitted by the function, $\rho = \rho_0 + \rho_1 T^\alpha$, with α ~ 1.9. The value determined from our experimental data is close to the above calculated one. This implies that the resistivity of the film from 40 K to 200 K is mainly attributed to the $MgB_2$ electron-phonon scattering. This further indicates that there is no excessive Mg in the present sample. Otherwise, the value of α would become larger due to the existence of excessive Mg in the film [15].

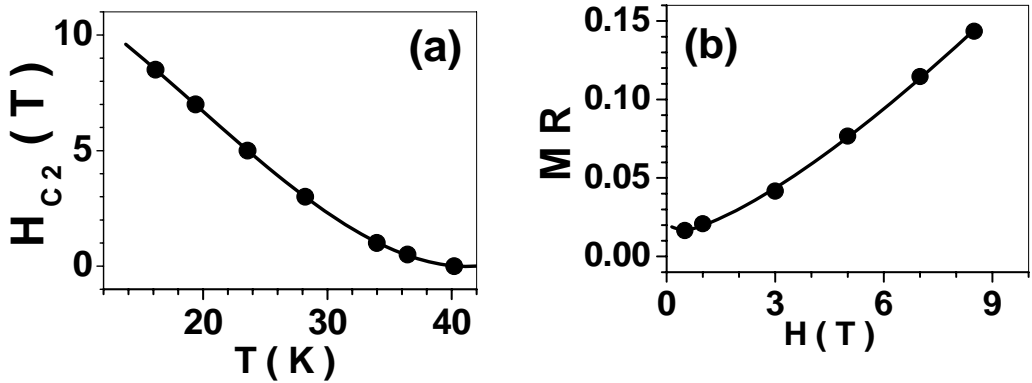

**Fig 3** (a) Upper critical field, $H_{C2}(T)$, at various temperatures. The points are fitted by the polynomial, $H_{C2}(T) = H_{C2}(0) + B_1T + B_2T^2 + B_3T^3$. (b) MR at 40 K. The points are fitted by the power law $\Delta\rho/\rho_0 = a_0 + a_1H^\beta$, in which $a_0$, $a_1$ and $\beta$ are the fitting parameters.

In the MR measurement shown in Fig. 2(b), the transition temperatures under various applied fields are determined by the criteria of 10 % below the normal state resistivity at 40 K. The points are plotted in Fig 3(a), and fitted with a polynomial, $H_{c2}(T) = H_{c2}(0) + B_1T + B_2T^2 + B_3T^3$. The extrapolation to zero temperature gives the upper critical field at $T = 0$ K, $H_{C2}(0) = 13.7$ T. Figure 3(a) shows that $H_{C2}(T)$ is curved upwards near $T_C$. This feature qualitatively agrees with the result of S.L. Bud'ko, etc. [16]. According to Gurevich on the two-band nature of $MgB_2$, if the σ band is dirtier than the π band, $H_{C2}(T)$ would be curved upwards near $T_C$; whereas, if the σ band is cleaner than the π band, the curve of $H_{C2}(T)$ near $T_C$ is linear with temperature and is curved upwards only near zero temperature [17]. The result plotted in Fig. 3(a) shows an up-curving feature near $T_C$ and a linear property at low temperature. This indicates that our $MgB_2$ film has a dirtier σ band than the π band. Note that $H_{C2}(0)$ of the present sample is not very high in comparison with the one of 28 T obtained in the $MgB_2$ film electroplated on the stainless steel substrate [18]. The MR at 40 K is determined directly from the $\rho$-$T$ measurements under various applied fields according to the equation, $MR(H) = \Delta\rho/\rho_0$, in which $\Delta\rho = \rho(H) - \rho_0$, and $\rho_0$ is the zero field resistivity. The result is plotted in Fig. 3(b) and fitted with the power law, $\Delta\rho/\rho_0 \propto H^\beta$. The exponent β is determined as 1.34, close to the value, 1.4, of dense $MgB_2$ wire [16]. It is noteworthy that MR of the present sample at $H = 5$ T is about 7.7 %. According to X.H.Chen et al [15], MR at $T = 50$ K and $H = 5$ T can be expressed by the scaling law as $\Delta\rho/\rho_0 = 0.04(RRR)^{2.2}$ %. Although we take the MR value at $T = 40$ K, we get the value of 7.8 % by applying this formula. This agrees very well with the scaling law proposed by X.H. Chen et.al.

The hysteresis measurement has been performed, using MPMSXL-7 magnetometer, at $T = 5$ K with the applied field varying from 4 T to - 4 T. The result is shown in the inset of Fig 4. The critical current density in Fig. 4, is estimated according to the Bean model using $J_C = 30 \Delta M/r$ [3], where $\Delta M$ is the height of the $M$-$H$ loop. Here we use $r = 3$ mm, which corresponds the radius calculated from the total area of the sample size according to $\pi r^2 = 10 \times 3$ mm$^2$. The critical current density at $T = 5$ K and zero field is about $5 \times 10^{10}$ A/m$^2$.

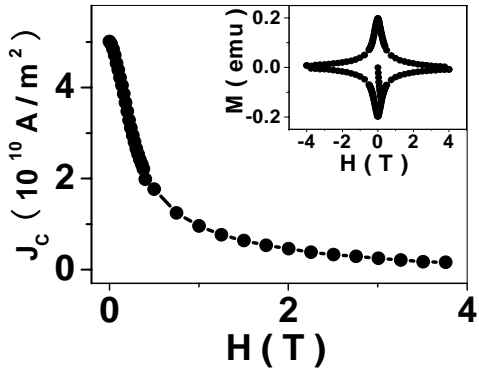

Fig. 4  Critical current density at 5 K estimated according to the Bean model. The inset shows the hysteresis loop measured at 5 K with the applied field ranging from 4 T to -4 T.

## 4  Conclusion

We have successfully synthesized thick $MgB_2$ film on the sapphire substrate by the method of HPCVD. The thickness of the film is about 1.3 μm. The transport measurement demonstrates that the thick $MgB_2$ film has a transition temperature at 40.2 K, with a sharp transition width of 0.15 K. By applying the Bean model, the critical superconducting current density, $J_C$, at $T$ = 5 K is about $5 \times 10^{10}$ A/m$^2$. The electron mean free path of the film at 40 K is about 10 nm, which is longer than the coherence length in the bulk, ~ 5 nm. It indicates that it is in the clean limit. MR of the film under $H$ = 8.5 T is 14 %. The large MR effect also suggests that the film is clean. Furthermore, the high residual resistivity ratio of 11 has been a sign of a clean sample as well. The upper critical field at $T$ = 0 K, $H_{c2}$(0), has been determined by extrapolation as 13.7 T. The present work has demonstrated that the HPCVD is an effective technique to fabricate high quality thick MgB$_2$ film, which is an important pre-stage to make superconducting tape. It is, therefore, a process with large application potential in the superconducting industry.

**Acknowledgement** This research is a part project of the Department of Physics of Peking University. Also, the project is supported by the National Natural Science Foundation of China under contract No. 50572001, and by the Special Foundation for State Major Basic Research Program of China (Grant No. 2006CB601007 ).